\documentclass[preprint,showpacs,preprintnumbers,amsmath,amssymb]{revtex4}

\usepackage{graphicx}
\usepackage{dcolumn}
\usepackage{bm}

\renewcommand{\bar}[1]{\overline{#1}}

\begin{document}

\preprint{~~}

\title{Principle of balance and the sea content of the proton}

\author{Yong-Jun Zhang} \email{zyj@pubms.pku.edu.cn}
\affiliation{Department of Physics, Peking University, Beijing
100871, China\\
Department of Basic Science, Liaoning Technical University, Fuxin,
Liaoning 123000, China}

\author{Wei-Zhen Deng} \email{dwz@pku.edu.cn}
\affiliation{Department of Physics, Peking University, Beijing
100871, China}

\author{Bo-Qiang Ma}
\email{mabq@phy.pku.edu.cn}\altaffiliation{corresponding author.}
\affiliation{ CCAST (World Laboratory), P.O.~Box 8730, Beijing
100080, China \\ Department of Physics, Peking University, Beijing
100871, China \footnote{Mailing address}}

\date{\today}

\begin{abstract}
In this study, the proton is taken as an ensemble of quark-gluon
Fock states. Using the principle of balance that every Fock state
should be balanced with all of the nearby Fock states (denoted as
the balance model), instead of the principle of detailed balance
that any two nearby Fock states should be balanced with each other
(denoted as the detailed balance model), the probabilities of
finding every Fock state of the proton are obtained. The balance
model can be taken as a revised version of the detailed balance
model, which can give an excellent description of the light flavor
sea asymmetry (i.e., $\bar{u}\not= \bar{d}$) without any
parameter. In case of $g\Leftrightarrow gg$ sub-processes not
considered, the balance model and the detailed balance model give
the same results. In case of $g\Leftrightarrow gg$ sub-processes
considered, there is about 10 percent difference between the
results of these models. We also calculate the strange content of
the proton using the balance model under the equal probability
assumption.
\end{abstract}



\maketitle

\section{Introduction}

One of the goals of hadronic physics is to describe the hadrons in
terms of their fundamental quark and gluon degrees of freedom from
basic principles. The structure of hadrons has been found to be
rather complicated due to the non-perturbative and relativistic
nature of the quark and gluon motions inside the hadrons. The sea
of hadrons plays a particular role as a source for the
complication. There have been many surprising discoveries
concerning the structure of the nucleon, and the sea content of
the nucleon has been found to be rather plentiful than natively
expected. For example, it was generally assumed that there should
be a symmetry between the light flavor $u$ and $d$ sea quarks
inside the proton. However, a surprisingly large asymmetry between
the $\bar{u}$ and $\bar{d}$ quark distributions of the proton has
been observed from experiments of both deep inelastic scattering
and Drell-Yan processes
\cite{FlavorAsymmetry,NMC91,NA51,HERMES,E866a,E866}. The strange
content of the proton sea is also found to be non-trivial in the
spin contribution \cite{Bro88} as well as in the quark-antiquark
content \cite{Bro96}.

Many theoretical attempts have been made to understand the sea
flavor asymmetry \cite{FlavorAsymmetry}, and it is believed that
the mesons inside the nucleon can account for such asymmetry.
Recently, there is a new attempt to understand the sea flavor
asymmetry of the proton from a pure statistical consideration in a
detailed balance model \cite{Zhang1,spintalk}. The idea is rather
simple and perspicuous: while the sea quark-antiquark $u\bar{u}$
and $d\bar{d}$ pairs can be produced by gluon splitting with equal
probabilities, the reverse processes of the annihilation of the
antiquarks with their quark partners into gluons are not flavor
symmetric due to the net excess of $u$ quarks over $d$ quarks. As
a consequence, the $\bar{u}$ quarks have large probability to
annihilate with the $u$ quarks than that of the $\bar{d}$ quarks,
and this brings an excess of $\bar{d}$ over $\bar{u}$ inside the
proton. Taking the proton as an ensemble of a complete set of
quark-gluon Fock states, and using the principle of detailed
balance that any two nearby Fock states should be balanced with
each other, one can obtain the probabilities of finding every Fock
state of the proton. Thus one can calculate the quark and gluon
content of the nucleon without any parameter from a pure
statistical consideration. It is interesting that the model gives
a sea flavor $\bar{u}$ and $\bar{d}$ asymmetry as $\bar{d}-\bar{u}
\sim 0.124$, which agrees surprisingly with the experimental data
$\bar{d}-\bar{u} = 0.118 \pm 0.012$. A further numerical
calculation \cite{Zhang2} also reproduced the explicit $x$
dependence of $\bar{d}(x)-\bar{u}(x)$, in good agreement with the
experimental data.

The purpose of this paper is to re-exam the detailed balance model
with more careful and general considerations. It will be shown
that the requirement of detailed balance, that any two nearby Fock
states should be balanced with each other, is an over-strong
constraint. It leads to inconsistency in cases if more channels of
sub-processes are introduced, as will be shown explicitly. We are
thus forced to start from a more general consideration which only
requires that every Fock state should be balanced with all of the
nearby Fock states, and then we construct the balance model as a
revised version of the detailed balance model. It will be shown
that in case of $g\Leftrightarrow gg$ sub-processes not
considered, the balance model and the detailed balance model give
the same results. In case of $g\Leftrightarrow gg$ sub-processes
considered, there is a difference of about 10 percent between the
results of the two models. We also calculate the strange content
of the proton using the balance model under the equal probability
assumption.

\section{General Principle of Balance instead of Detailed Balance}

In the detailed balance model \cite{Zhang1,spintalk}, the proton
state is expended by a complete set of quark-gluon Fock states as
\begin{equation}
| p \rangle=\sum_{i,j,k} c_{i,j,k}|uud,i,j,k\rangle,
\label{FockStates}
\end{equation}
where $i$ is the number of quark-antiquark $u\bar{u}$ pairs, $j$
is the number of quark-antiquark $d \bar{d}$ pairs, and $k$ is the
number of gluons. Then the probability of finding the proton in
the Fock state $\left|uud,i,j,k\right\rangle$ is
\begin{equation}
\rho_{i,j,k}=\left| c_{i,j,k}\right|^2,
\end{equation}
where $\rho_{i,j,k}$ satisfies the normalization condition,
\begin{equation}
\sum_{i,j,k} \rho_{i,j,k}=1. \label{unit}
\end{equation}
Then using the detailed balance principle, and with sub-processes
$q\Leftrightarrow q g$ and $g \Leftrightarrow q\bar{q}$
considered, one can calculate all $\rho_{i,j,k}$ explicitly
\cite{Zhang1,spintalk}. The detailed balance principle means that
any two nearby quark-gluon Fock states should balance with each
other. However, when the detailed balance principle is used to the
case with the sub-processes $g\Leftrightarrow gg$ also considered,
there is an inconsistency which is not easy to be noticed. For
example, there are two ways to calculate
${\rho_{uud\bar{u}ug}}$/${\rho_{uudg}}$. One way is
\begin{eqnarray}
|uudg\rangle{\footnotesize\begin{array}{c}3+1\\
{\Large\rightleftharpoons}
\\ 2\times3+1 \end{array}}|uudgg\rangle
{\footnotesize\begin{array}{c}2\\
{\Large\rightleftharpoons}
\\ 1\times3 \end{array}} |uud\bar{u}ug\rangle,
\end{eqnarray}
which leads to
\begin{eqnarray}
\rho_{uudgg}=\frac{4}{7}\rho_{uudg},\
\rho_{uud\bar{u}ug}=\frac{2}{3}\rho_{uudgg},\
\frac{\rho_{uud\bar{u}ug}}{\rho_{uudg}}=\frac{8}{21}. \ \
\end{eqnarray}
Another way is
\begin{eqnarray}
|uudg\rangle{\footnotesize\begin{array}{c}1\\
{\Large\rightleftharpoons}
\\ 1\times3 \end{array}}|uud\bar{u}u\rangle
{\footnotesize\begin{array}{c}5\\
{\Large\rightleftharpoons}
\\ 1\times5 \end{array}} |uud\bar{u}ug\rangle,
\end{eqnarray}
which leads to
\begin{eqnarray}
\rho_{uud\bar{u}u}=\frac{1}{3}\rho_{uudg},\
\rho_{uud\bar{u}ug}=\rho_{uudgg},\
\frac{\rho_{uud\bar{u}ug}}{\rho_{uudg}}&=&\frac{1}{3}. \ \
\end{eqnarray}
One way gives the result $\frac{8}{21}$, while another way gives
the result $\frac{1}{3}$. This inconsistency forces us to adopt a
more general principle named the balance principle instead the
detailed balance principle.

The basic property of the ensemble assumption of the proton is
that the probability to find the proton in any Fock state should
not change during the time. Both the detailed balance principle
and balance principle can give this property of ensemble
assumption.

The detailed balance demands that any two nearby Fock states $A$
and $B$ balance with each other \cite{spintalk},
 \begin{eqnarray}
\rho_A R_{A\rightarrow B}=\rho_B R_{B\rightarrow
A},\label{detailed-balance}
\end{eqnarray}
where $\rho_A$ is the probability of finding the proton in state
of $A$, $\rho_B$ is the probability of finding the proton in state
$B$, $R_{A\rightarrow B}$ is the transition probability of $A$ to
$B$, and $R_{B\rightarrow A}$ is the transition probability of $B$
to $A$.

The balance principle demands that the ``go out"  probability just
balances the ``come in" probability for any Fock state. For a Fock
state $A$, the ``go out" probability is
\begin{eqnarray}
\sum_{B} \rho_A R_{A\rightarrow B},
\end{eqnarray}
 where $B$ is a completed set of all nearby Fock states other than
 $A$,
 while the ``come in" probability is
\begin{eqnarray}
\sum_{B} \rho_B R_{B\rightarrow A}.
\end{eqnarray}
 So the balance principle can be written as
 \begin{eqnarray}
\sum_{B} \rho_A R_{A\rightarrow B}=\sum_{B}\rho_BR_{B\rightarrow
A}. \label{balance}
\end{eqnarray}
 From the formulae (\ref{detailed-balance}) and (\ref{balance}), it
is easy to find that the detailed balance principle
(\ref{detailed-balance}) gives more strong constraint than the
balance principle (\ref{balance}). That is why the detailed
balance principle leads to the inconsistency discussed above while
the balance principle does not. In similar to the detailed balance
principle, the balance principle can also determine all
$\rho_{i,j,k}$, but the situation becomes a little more
complicated and solving an equation set with a large number of
equations is needed.

\section{The Balance Model with no $g\Leftrightarrow gg$
sub-Processes Considered}

In this section, we only consider the transition processes
involving the sub-processes $q \Leftrightarrow q g$ and $g
\Leftrightarrow q \bar{q}$, with the sub-processes
$g\Leftrightarrow gg$ not considered at first. When applying the
balance principle to Fock state $|uud\rangle$, we have the ``go
out" process as
\begin{eqnarray}
|uud\rangle{\footnotesize\begin{array}{c}3\\
{\Large\Rightarrow}\\{}
\end{array}}|uudg\rangle,
\end{eqnarray}
and the ``come in" process as
\begin{eqnarray}
|uud\rangle{\footnotesize\begin{array}{c}3\\
{\Large\Leftarrow}\\{}\end{array}}|uudg\rangle.
\end{eqnarray}
According to the formula (\ref{balance}), we have the equation
\begin{eqnarray}
3\rho_{uud}=3\rho_{uudg} \label{000}.
\end{eqnarray}
When using the balance principle to the Fock state $|uudg\rangle$,
we have the ``go out" processes as
\begin{eqnarray}
|uudg\rangle &{\footnotesize\begin{array}{c}3\times 1\\
{\Large\Rightarrow}\\{}\end{array}} &|uud\rangle,\\
|uudg\rangle &{\footnotesize\begin{array}{c}3\\
{\Large\Rightarrow}\\{}\end{array}} &|uudgg\rangle,\\
|uudg\rangle &{\footnotesize\begin{array}{c}1\\
{\Large\Rightarrow}\\{}\end{array}} &|uud\bar{u}u\rangle,\\
|uudg\rangle &{\footnotesize\begin{array}{c}1\\
{\Large\Rightarrow}\\{}\end{array}} &|uud\bar{d}d\rangle,
\end{eqnarray}
and the ``come in" processes as
\begin{eqnarray}
|uudg\rangle &{\footnotesize\begin{array}{c}3\\
{\Large\Leftarrow}\\{}\end{array}} &|uud\rangle,\\
|uudg\rangle &{\footnotesize\begin{array}{c}3\times 2\\
{\Large\Leftarrow}\\{}\end{array}} &|uudgg\rangle,\\
|uudg\rangle &{\footnotesize\begin{array}{c}1\times 3\\
{\Large\Leftarrow}\\{}\end{array}} &|uud\bar{u}u\rangle,\\
|uudg\rangle &{\footnotesize\begin{array}{c}1\times 2\\
{\Large\Leftarrow}\\{}\end{array}} &|uud\bar{d}d\rangle.
\end{eqnarray}
 From the formula (\ref{balance}), we have the equation
\begin{eqnarray}
&\left(3\times 1+3+1+1\right)\rho_{uudg} =  3\rho_{uud}
&+(3\times 2) \rho_{uudgg}+(1\times
3)\rho_{uud\bar{u}u}+(1\times 2)\rho_{uud\bar{d}d} ~. \ \
\end{eqnarray}
When using the balance principle to a general Fock state
$|uud,i,j,k\rangle$, we have the ``go out" processes as
\begin{eqnarray}
|uud,i,j,k\rangle &{\footnotesize\begin{array}{c}(3+2i+2j)k\\
{\Large\Rightarrow}\\{}\end{array}} &|uud,i,j,k-1\rangle,\label{out-noggg}\\
|uud,i,j,k\rangle &{\footnotesize\begin{array}{c}3+2i+2j\\
{\Large\Rightarrow}\\{}\end{array}} &|uud,i,j,k+1\rangle,\\
|uud,i,j,k\rangle &{\footnotesize\begin{array}{c}k\\
{\Large\Rightarrow}\\{}\end{array}} &|uud,i+1,j,k-1\rangle,\\
|uud,i,j,k\rangle &{\footnotesize\begin{array}{c}i(i+2)\\
{\Large\Rightarrow}\\{}\end{array}} &|uud,i-1,j,k+1\rangle,\\
|uud,i,j,k\rangle &{\footnotesize\begin{array}{c}k\\
{\Large\Rightarrow}\\{}\end{array}} &|uud,i,j+1,k-1\rangle,\\
|uud,i,j,k\rangle &{\footnotesize\begin{array}{c}j(j+1)\\
{\Large\Rightarrow}\\{}\end{array}} &|uud,i,j-1,k+1\rangle,
\end{eqnarray}
and the ``come in" processes as
\begin{eqnarray}
|uud,i,j,k\rangle &{\footnotesize\begin{array}{c}3+2i+2j \\
{\Large\Leftarrow}\\{}\end{array}} &|uud,i,j,k-1\rangle,\\
|uud,i,j,k\rangle &{\footnotesize\begin{array}{c}(3+2i+2j)(k+1)\\
{\Large\Leftarrow}\\{}\end{array}} &|uud,i,j,k+1\rangle,\\
|uud,i,j,k\rangle &{\footnotesize\begin{array}{c}(i+1)(i+3)\\
{\Large\Leftarrow}\\{}\end{array}} &|uud,i+1,j,k-1\rangle,\\
|uud,i,j,k\rangle &{\footnotesize\begin{array}{c}k+1\\
{\Large\Leftarrow}\\{}\end{array}} &|uud,i-1,j,k+1\rangle,\\
|uud,i,j,k\rangle &{\footnotesize\begin{array}{c}(j+1)(j+2)\\
{\Large\Leftarrow}\\{}\end{array}} &|uud,i,j+1,k-1\rangle,\\
|uud,i,j,k\rangle &{\footnotesize\begin{array}{c}k+1\\
{\Large\Leftarrow}\\{}\end{array}} &|uud,i,j-1,k+1\rangle. \ \
\label{in-noggg}
\end{eqnarray}
Then from the formula (\ref{balance}), we have the equation
\begin{eqnarray}
&& \left[(3+2i+2j)k+(3+2i+2j)+k+i(i+2)+k+j(j+1)\right]
\rho_{i,j,k}\nonumber\\
=&&(3+2i+2j)\rho_{i,j,k-1}+(3+2i+2j)(k+1)\rho_{i,j,k+1}
+(i+1)(i+3)\rho_{i+1,j,k-1}\nonumber \\
&&+(k+1)\rho_{i-1,j,k+1}+(j+1)(j+2)\rho_{i,j+1,k-1}+(k+1)\rho_{i,j-1,k+1}.
\label{equation-noggg}
\end{eqnarray}
When one applies the equation (\ref{equation-noggg}) to a certain
Fock state, he should make sure that the corresponding Fock states
in processes (\ref{out-noggg}-\ref{in-noggg}) exist, otherwise he
should delete the corresponding terms. For example, when one
applies the equation (\ref{equation-noggg}) to the Fock state
$|uud\rangle$, only one corresponding Fock state in processes
(\ref{out-noggg}-\ref{in-noggg}) exists, i.e., it is
$|uudg\rangle$. So the equation (\ref{equation-noggg}) leads to
equation (\ref{000}).

For every Fock state $|uud,i,j,k\rangle$, there is an equation
according to formula (\ref{equation-noggg}). If we can construct
an equation set using these equations, then $\rho_{i,j,k}$ can be
solved. However, there is an infinite number of Fock states that
lead to an infinite number of equations.  We must choose the upper
limits of $i$,$j$ and $k$ to select a finite number of Fock
states. Then by increasing the upper limits, we can approach the
values for $\rho_{i,j,k}$ because $\rho_{i,j,k}$ decreases quickly
in a manner somewhat like $\frac{1}{i!j!k!}$ as $i$, $j$ and $k$
increasing. When choosing the upper limits of $i$ to $N_i$, $j$ to
$N_j$ and $k$ to $N_k$, we get
\begin{equation}
\left| p
\right\rangle=\sum_{k=0}^{N_k}\sum_{j=0}^{N_j}\sum_{i=0}^{N_i}
c_{i,j,k}\left| uud, i,j,k \right\rangle. \label{cut}
\end{equation}
There are $N=(N_i+1)(N_j+1)(N_k+1)$ Fock states selected. So there
are $N$ numbers of $\rho_{i,j,k}$ to be solved. There are $N$
equations when using formula (\ref{equation-noggg}) to these $N$
Fock states. However, these $N$ equations are not independent. In
order to show this fact, we write the formula (\ref{balance}) in
the new form by including these upper limits as
\begin{equation}
\sum_{m^{\prime}=1}^N \rho_{A_m}R_{A_m\rightarrow A_{m^{\prime}}}=
\sum_{m^{\prime}=1}^N
\rho_{A_{m^{\prime}}}R_{A_{m^{\prime}}\rightarrow A_{m}},
\label{prove}
\end{equation}
where $A_m$ and $A_{m^{\prime}}$ are any two Fock states of those
$N$ Fock states selected by formula (\ref{cut}) and we let
$R_{A_m\rightarrow A_{m}}=0$ which gives no any physical effect
but is convenience for calculation in mathematics. There are $N$
such equations as formula (\ref{prove}) for $N$ selected Fock
states. Summing over all these equations, we get
\begin{equation}
\sum_{m=1}^N\sum_{m^{\prime}=1}^N \rho_{A_m}R_{A_m\rightarrow
A_{m^{\prime}}}= \sum_{m=1}^N\sum_{m^{\prime}=1}^N
\rho_{A_{m^{\prime}}}R_{A_{m^{\prime}}\rightarrow A_{m}}.
\label{prove-sum}
\end{equation}
On the other side, from the pure mathematical point, we have the
relation of
\begin{equation}
\sum_{m=1}^N\sum_{m^{\prime}=1}^N \rho_{A_m}R_{A_m\rightarrow
A_{m^{\prime}}}
\begin{array}{c}{\rm exchange\ }m{\rm\ and\ }m^{\prime}\\
 \equiv\\
\end{array}
 \sum_{m^{\prime}=1}^N\sum_{m=1}^N
\rho_{A_{m^{\prime}}}R_{A_{m^{\prime}}\rightarrow A_{m}} =
 \sum_{m=1}^N\sum_{m^{\prime}=1}^N \rho_{A_{m^{\prime}}}
 R_{A_{m^{\prime}}\rightarrow A_{m}},
\end{equation}
which is the same form as relation (\ref{prove-sum}). So the
relation (\ref{prove-sum}) does not depend on the existence of
equation (\ref{prove}) and it is always right in mathematics. It
means that the relation (\ref{prove-sum}) is a constraint of these
$N$ equations so that there are only $N-1$ independent equations
left. With the normalization condition (\ref{unit}) included,
there are just $N$ equations to construct a complete equation set
for $N$ variables ($\rho_{i,j,k}$). Thus this equation set can be
solved.

By solving this equation set numerically, we find that the
balance model and detailed balance model give the same results in
case of sub-processes $g\Leftrightarrow gg$ not considered. From a
strict sense, by substituting the solution of the detailed
balanced model, i.e., Eq.~(26) of Ref.~\cite{Zhang1},
\begin{equation}
\rho_{i,j,k}=\frac{2}{i!(i+2)!j!(j+1)!k!}\rho_{0,0,0},\label{aijk}
\end{equation}
for the corresponding terms in equation (\ref{equation-noggg}), we
find that (\ref{equation-noggg}) is exactly satisfied. This proves
that the balanced model has the exact solution as that of the
detailed balance model.

\section{The Balance Model with $g\Leftrightarrow gg$
sub-Processes Considered}

In this section, the processes involving sub-processes
$g\Leftrightarrow gg$ are also considered. When using the balance
principle to Fock state $|uud\rangle$, we have the ``go out"
process as
\begin{eqnarray}
|uud\rangle{\footnotesize\begin{array}{c}3\\
{\Large\Rightarrow}\\{}\end{array}}|uudg\rangle,
\end{eqnarray}
and the ``come in" process as
\begin{eqnarray}
|uud\rangle{\footnotesize\begin{array}{c}3\\
{\Large\Leftarrow}\\{}\end{array}}|uudg\rangle.
\end{eqnarray}
According to the formula (\ref{balance}), we have the equation
\begin{eqnarray}
3\rho_{uud}=3\rho_{uudg}.
\end{eqnarray}
When using the balance principle to the Fock state $|uudg\rangle$,
we have the ``go out" processes as
\begin{eqnarray}
|uudg\rangle &{\footnotesize\begin{array}{c}3\times 1\\
{\Large\Rightarrow}\\{}\end{array}} &|uud\rangle,\\
|uudg\rangle &{\footnotesize\begin{array}{c}3+1\\
{\Large\Rightarrow}\\{}\end{array}} &|uudgg\rangle,\\
|uudg\rangle &{\footnotesize\begin{array}{c}1\\
{\Large\Rightarrow}\\{}\end{array}} &|uud\bar{u}u\rangle,\\
|uudg\rangle &{\footnotesize\begin{array}{c}1\\
{\Large\Rightarrow}\\{}\end{array}} &|uud\bar{d}d\rangle,
\end{eqnarray}
and the ``come in" processes as
\begin{eqnarray}
|uudg\rangle &{\footnotesize\begin{array}{c}3\\
{\Large\Leftarrow}\\{}\end{array}} &|uud\rangle,\\
|uudg\rangle &{\footnotesize\begin{array}{c}3\times 2+1\\
{\Large\Leftarrow}\\{}\end{array}} &|uudgg\rangle,\\
|uudg\rangle &{\footnotesize\begin{array}{c}1\times 3\\
{\Large\Leftarrow}\\{}\end{array}} &|uud\bar{u}u\rangle,\\
|uudg\rangle &{\footnotesize\begin{array}{c}1\times 2\\
{\Large\Leftarrow}\\{}\end{array}} &|uud\bar{d}d\rangle.
\end{eqnarray}
 From the formula (\ref{balance}), we have the equation
\begin{eqnarray}
\left(3\times 1+(3+1)+1+1\right)\rho_{uudg}=3\rho_{uud}+(3\times
2+1) \rho_{uudgg}+(1\times 3)\rho_{uud\bar{u}u}+(1\times
2)\rho_{uud\bar{d}d}. \nonumber
\end{eqnarray}
When using the balance principle to a general Fock state
$|uud,i,j,k\rangle$, we have the ``go out" processes as
\begin{eqnarray}
|uud,i,j,k\rangle &{\footnotesize\begin{array}{c}(3+2i+2j)k+C_k^2\\
{\Large\Rightarrow}\\{}\end{array}} &|uud,i,j,k-1\rangle,\\
|uud,i,j,k\rangle &{\footnotesize\begin{array}{c}3+2i+2j+k\\
{\Large\Rightarrow}\\{}\end{array}} &|uud,i,j,k+1\rangle,\\
|uud,i,j,k\rangle &{\footnotesize\begin{array}{c}k\\
{\Large\Rightarrow}\\{}\end{array}} &|uud,i+1,j,k-1\rangle,\\
|uud,i,j,k\rangle &{\footnotesize\begin{array}{c}i(i+2)\\
{\Large\Rightarrow}\\{}\end{array}} &|uud,i-1,j,k+1\rangle,\\
|uud,i,j,k\rangle &{\footnotesize\begin{array}{c}k\\
{\Large\Rightarrow}\\{}\end{array}} &|uud,i,j+1,k-1\rangle,\\
|uud,i,j,k\rangle &{\footnotesize\begin{array}{c}j(j+1)\\
{\Large\Rightarrow}\\{}\end{array}} &|uud,i,j-1,k+1\rangle,
\end{eqnarray}
and the ``come in" processes as
\begin{eqnarray}
|uud,i,j,k\rangle &{\footnotesize\begin{array}{c}3+2i+2j+k-1\\
{\Large\Leftarrow}\\{}\end{array}} &|uud,i,j,k-1\rangle,\\
|uud,i,j,k\rangle &{\footnotesize\begin{array}{c}(3+2i+2j)(k+1)+C_{k+1}^2\\
{\Large\Leftarrow}\\{}\end{array}} &|uud,i,j,k+1\rangle,\\
|uud,i,j,k\rangle &{\footnotesize\begin{array}{c}(i+1)(i+3)\\
{\Large\Leftarrow}\\{}\end{array}} &|uud,i+1,j,k-1\rangle,\\
|uud,i,j,k\rangle &{\footnotesize\begin{array}{c}k+1\\
{\Large\Leftarrow}\\{}\end{array}} &|uud,i-1,j,k+1\rangle,\\
|uud,i,j,k\rangle &{\footnotesize\begin{array}{c}(j+1)(j+2)\\
{\Large\Leftarrow}\\{}\end{array}} &|uud,i,j+1,k-1\rangle,\\
|uud,i,j,k\rangle &{\footnotesize\begin{array}{c}k+1\\
{\Large\Leftarrow}\\{}\end{array}} &|uud,i,j-1,k+1\rangle.
\end{eqnarray}
 From the formula (\ref{balance}), we have the equation
\begin{eqnarray}
&&\left\{
\left[(3+2i+2j)k+C_k^2\right]+(3+2i+2j+k)+k+i(i+2)+k+j(j+1)\right\}
\rho_{i,j,k}\nonumber\\
=&&(3+2i+2j+k-1)\rho_{i,j,k-1}+\left[(3+2i+2j)(k+1)+C_{k+1}^2\right]\rho_{i,j,k+1}
+(i+1)(i+3)\rho_{i+1,j,k-1}\nonumber \\
&&+(k+1)\rho_{i-1,j,k+1}+(j+1)(j+2)\rho_{i,j+1,k-1}+(k+1)\rho_{i,j-1,k+1},
\label{equationggg}
\end{eqnarray}
where $C_k=k(k-1)/2$. An equation set for $\rho_{i,j,k}$ can be
constructed by using the above equation after choosing the upper
limits on $i$, $j$ and $k$. We can solve the equation set by
increasing the upper limits to a situation that the results do not
sensitive to the upper limits further. Thus $\rho_{i,j,k}$ can be
obtained and the results are shown in Table \ref{table}. Data of
Table \ref{table} show that the balance model give different
results compared with that of the detailed balance model. From
data of Table \ref{table} we have
\begin{eqnarray}
&\bar{u}=\sum_{i,j,k}i\rho_{i,j,k}=0.337,\\ &\bar{d}=\sum_{i,j,k}
j\rho_{i,j,k}=0.470,\\ & g=\sum_{i,j,k} k\rho_{i,j,k}=1.099,\\ &
\bar{d}-\bar{u}=0.133.
 \end{eqnarray}
We notice that $\bar{d}-\bar{u}=0.133$ for the light-flavor sea
asymmetry in comparison with the result of detailed balance
 model  $\bar{d}-\bar{u}=0.123$ \cite{Zhang1}, while
 the FNAL/NuSea experiment is $\bar{d}-\bar{u}=0.118\pm
 0.012$  \cite{E866}. The balance model is different from the detailed balance
 model by about 10\%, with both are compatible with the experimental
 result.

\section{Strange Quark Content and Gluon Energy Distribution}

It has been known that the strange sea is non-trial inside the
proton, and a phenomenological analysis suggests that the number
of $s \bar{s}$ pairs inside a proton is about $0.05$ \cite{Bro96}.
Here we extend the balance model to include the strange
quark-antiquark pairs, and write the proton state as
\begin{equation}
| p \rangle=\sum_{i,j,k,l} c_{i,j,k,l}|uud,i,j,k,l\rangle,
\label{FockStates-strange}
\end{equation}
where $l$ is the number of quark-antiquark $s\bar{s}$ pairs. Then
the probability of finding the proton in the Fock state
$\left|uud,i,j,k,l\right\rangle$ is
\begin{equation}
\rho_{i,j,k,l}=\left| c_{i,j,k,l}\right|^2,
\end{equation}
where $\rho_{i,j,k,l}$ satisfies the normalization condition,
\begin{equation}
\sum_{i,j,k,l} \rho_{i,j,k,l}=1. \label{unit-strange}
\end{equation}
The $s\bar{s}$ pairs are generated from gluons by the
sub-processes $g\Rightarrow s\bar{s}$. However the mass of the
strange quark $M_s$ is big and should not be neglected as for the
light-flavor quarks. So only those gluons with energy large than 2
times of the strange quark mass, $\varepsilon_g>2M_s$, can take
part in the process of splitting into strange quark-antiquark
pairs by the constraint of energy conservation. So we should know
the gluon energy distribution before the calculation of the
strange content of the proton.

An equal probability assumption is presented in paper
\cite{Zhang2}, assuming that for a $n$-parton Fock state
$|uud,i,j,k\rangle$, any parton's energy configuration
($E_1,\cdots,E_n$) has the same probability to appear,
\begin{eqnarray}
d\rho_n(p_1,\cdots,p_n)=\delta^4(P-\sum^n_{i=1} p_i)\prod^n_{i=1}
\frac{dE_id\Omega_i}{2(2\pi)^3}, \label{zoubs}
\end{eqnarray}
with $P$ the 4-momentum of the proton, and $p_1,\cdots,p_n$ the
4-momenta of $n$ partons. The formula (\ref{zoubs}) is equivalent
to a series of constraints,
\begin{eqnarray}
&f^n({E}_1,E_2,\cdots,E_n)\propto 1,\label{equal probability}\\
&E_1^2-p_{1x}^2-p_{1y}^2-p_{1z}^2=0,\label{onshell1}\\
&E_2^2-p_{2x}^2-p_{2y}^2-p_{2z}^2=0,\label{onshell2}\\
&\cdots,\nonumber\\
&E_n^2-p_{nx}^2-p_{ny}^2-p_{nz}^2=0,\label{onshell3}\\
&f_{E_1}(p_{1x}^2,p_{1y}^2,p_{1z}^2)\propto 1,\label{momentum1}\\
&f_{E_2}(p_{2x}^2,p_{2y}^2,p_{2z}^2)\propto 1,\label{momentum2}\\
&\cdots,\nonumber\\ &f_{E_n}(p_{nx}^2,p_{ny}^2,p_{nz}^2)\propto
1,\label{momentum3}\\
&p_{1x}+p_{2x}+\cdots+p_{nx}=0,\label{restframe1}\\
&p_{1y}+p_{2y}+\cdots+p_{ny}=0,\label{restframe2}\\
&p_{1z}+p_{2z}+\cdots+p_{nz}=0,\label{restframe3}\\
&E_1+E_2+\cdots+E_n=M_{\rm proton}=938 {\rm
MeV},\label{restframe4}
\end{eqnarray}
where (\ref{equal probability}) gives the principle of equal
probability, (\ref{onshell1})-(\ref{onshell3}) mean that the
partons are massless and on shell,
(\ref{momentum1})-(\ref{momentum3}) mean that the distribution
does not depend on the 3-momentum, and
(\ref{restframe1})-(\ref{restframe4}) mean that the
energy-momentum conservation constraint is satisfied and our
calculation is in the rest frame of proton.

If we only concern about the energy distribution, then from
constraint (\ref{equal probability}) and (\ref{restframe4}), we
may get it as,
\begin{eqnarray}
f^n(E_1,E_2,\cdots,E_n)=(n-1)!\delta(1-\frac{E_1}{M_{\rm proton}}
-\frac{E_2}{M_{\rm proton}}-\cdots-\frac{E_n}{M_{\rm
proton}}),\label{energy-d}
\end{eqnarray}
which satisfies the normalization condition.

Before we extend the formula (\ref{energy-d}) to include the
strange quark, a new concept named free energy needs to be
introduced to deal with the strange quark mass $M_s$. For a
$n$-parton Fock state $|uud,i,j,k,l\rangle$ where
$n=3+2i+2j+k+2l$, the parton's free energy is presented as $\
\varepsilon_1,\varepsilon_2,\cdots,\varepsilon_n$. Here the
parton's free energy denotes the energy of a parton that can be
used to produce other partons, and it is defined as the total
energy of one parton minus the mass of that parton,
$\varepsilon=E-M$. For the partons as $u$, $d$, and $g$, they are
almost massless so that we have
\begin{eqnarray}
\varepsilon=E.
\end{eqnarray}
For the strange quark $s$, we have
\begin{eqnarray}
\varepsilon=E-M_s \ .
\end{eqnarray}
Then the total free energy of the Fock state $|uud,i,j,k,l\rangle$
is
\begin{eqnarray}
\xi_f=\sum_m \varepsilon_m=\sum_m E_m-2lM_s=M_{\rm proton}-2lM_s,
\end{eqnarray}
where $M_{\rm proton}$ is the mass of the proton.

We define the dimensionless free energy as
\begin{eqnarray}
y_1=\frac{\varepsilon_1}{\xi_f},\\
y_2=\frac{\varepsilon_2}{\xi_f},\\
\cdots
\nonumber
\\
y_n=\frac{\varepsilon_n}{\xi_f},
\end{eqnarray}
which lead to
\begin{eqnarray}
\sum_m y_m=1.
\end{eqnarray}

Replacing the variable of energy $E$ by the variable of free
energy $\varepsilon$, we rewrite the energy distribution formula
(\ref{energy-d}) to a form as
\begin{eqnarray}
f^n(y_1,y_2,\cdots,y_n)=(n-1)!\delta(1-y_1-y_2-\cdots-y_n),
\end{eqnarray}
which can be used to deal with $s$ quarks as well as $u$ quarks,
$d$ quarks, and gluons.

Then for one parton in a $n$-parton Fock state, the free energy
distribution is
\begin{eqnarray}
f^n(y)=&&\int_0^1 dy_1 \cdots \int_0^1
dy_{n-1}f^n(y,y_1,\cdots,y_{n-1})\nonumber\\
=&&\int_0^1 dy_1\cdots \int_0^1 dy_{n-1}(n-1)! \delta(1-y-y_1-\cdots-y_{n-1})\nonumber\\
=&&(n-1)(1-y)^{n-2},\label{induction}
\end{eqnarray}
which can be proved using the following completed derivation.

It is easy to see
\begin{eqnarray}
&\int_0^1 dy_1 \delta (1-y-y_1)=1,
\\ &\int_0^1 dy_1 \int_0^1 dy_2
2\delta (1-y-y_1-y_2)=\int_0^{1-y}dy_12=2(1-y),
\end{eqnarray}
if
\begin{eqnarray}
\int_0^1 dy_1 \cdots \int_0^1 dy_{n-1}
{(n-1)!}\delta(1-y-y_1-\cdots-y_{n-1})=(n-1)(1-y)^{n-2},
\end{eqnarray}
then
\begin{eqnarray}
&&\int_0^1 dy_1 \cdots \int_0^1 dy_{n} n!
\delta(1-y-y_1-\cdots-y_{n-1}-y_{n})\\&=&\int_0^{1-y} dy_{n}
n(n-1)(1-y-y_{n})^{n-2}\\&=&n(1-y)^{n-1},
\end{eqnarray}
so formula (\ref{induction}) is proved.

Applying formula (\ref{induction}) to a gluon in a $n$-parton Fock
state $|uud,i,j,k,l\rangle$, we get the gluon's free energy
distribution
\begin{eqnarray}
f_g^n(y)=&&(n-1)(1-y)^{n-2}.
\end{eqnarray}
Then we introduce a parameter
\begin{eqnarray}
c_l=\frac{2M_s}{\xi_f}=\frac{2M_s}{M_{\rm proton}-2lM_s}.
\end{eqnarray}
Only those gluons satisfying $y>c_l$ have the energy $E>2M_s$, so
that they can split into $\bar{s}s$ pairs. Then for a gluon in a
$n$-parton Fock state $|uud,i,j,k,l\rangle$, the probability that
it can split to a $\bar{s}s$ pair is
\begin{eqnarray}
\int_{c_l}^1 dy (n-1)(1-y)^{n-2}=(1-c_l)^{n-1}.
\end{eqnarray}
Taking $(1-c_l)^{n-1}$ as a suppressing factor of generating
$s\bar{s}$ from a gluon, the content of strange quark of proton
can be calculated by using the balance model.

\subsection{Strange content with no $g\Leftrightarrow gg$
sub-processes considered}

We first consider the strange quark content of the proton with the
processes involving the $g\Leftrightarrow gg$ sub-processes not
considered. When using the balance principle to Fock state
$|uud\rangle$, we have the ``go out" process as
\begin{eqnarray}
|uud\rangle{\footnotesize\begin{array}{c}3\\
{\Large\Rightarrow}\\{}\end{array}}|uudg\rangle,
\end{eqnarray}
and the ``come in" process as
\begin{eqnarray}
|uud\rangle{\footnotesize\begin{array}{c}3\\
{\Large\Leftarrow}\\{}\end{array}}|uudg\rangle.
\end{eqnarray}
According to the formula (\ref{balance}), we have the equation
\begin{eqnarray}
3\rho_{uud}=3\rho_{uudg}.
\end{eqnarray}
When using the balance principle to Fock state $|uudg\rangle$, we
have the ``go out" processes as
\begin{eqnarray}
|uudg\rangle &{\footnotesize\begin{array}{c}3\times 1\\
{\Large\Rightarrow}\\{}\end{array}} &|uud\rangle,\\
|uudg\rangle &{\footnotesize\begin{array}{c}3\\
{\Large\Rightarrow}\\{}\end{array}} &|uudgg\rangle,\\
|uudg\rangle &{\footnotesize\begin{array}{c}1\\
{\Large\Rightarrow}\\{}\end{array}} &|uud\bar{u}u\rangle,\\
|uudg\rangle &{\footnotesize\begin{array}{c}1\\
{\Large\Rightarrow}\\{}\end{array}} &|uud\bar{d}d\rangle,\\
|uudg\rangle &{\footnotesize\begin{array}{c}1\times(1-c_0)^3\\
{\Large\Rightarrow}\\{}\end{array}} &|uud\bar{s}s\rangle,
\end{eqnarray}
and the ``come in" processes as
\begin{eqnarray}
|uudg\rangle &{\footnotesize\begin{array}{c}3\\
{\Large\Leftarrow}\\{}\end{array}} &|uud\rangle,\\
|uudg\rangle &{\footnotesize\begin{array}{c}3\times 2\\
{\Large\Leftarrow}\\{}\end{array}} &|uudgg\rangle,\\
|uudg\rangle &{\footnotesize\begin{array}{c}1\times 3\\
{\Large\Leftarrow}\\{}\end{array}} &|uud\bar{u}u\rangle,\\
|uudg\rangle &{\footnotesize\begin{array}{c}1\times 2\\
{\Large\Leftarrow}\\{}\end{array}} &|uud\bar{d}d\rangle,\\
|uudg\rangle &{\footnotesize\begin{array}{c}1\times 1\\
{\Large\Leftarrow}\\{}\end{array}} &|uud\bar{s}s\rangle.
\end{eqnarray}
Then we have the equation
\begin{eqnarray}
&&\left[3\times
1+3+1+1+1\times(1-c_0)^3\right]\rho_{uudg}\nonumber
\\ =&&3\rho_{uud}+(3\times 2) \rho_{uudgg}+(1\times
3)\rho_{uud\bar{u}u}+(1\times 2)\rho_{uud\bar{d}d}+(1\times
1)\rho_{uud\bar{s}s}
\end{eqnarray}

Applying the balance principle to a general Fock state
$|uud,i,j,k,l\rangle$, we have the ``go out" processes as
\begin{eqnarray}
|uud,i,j,k,l\rangle &{\footnotesize\begin{array}{c}(3+2i+2j+2l)k\\
{\Large\Rightarrow}\\{}\end{array}} &|uud,i,j,k-1,l\rangle,\\
|uud,i,j,k,l\rangle &{\footnotesize\begin{array}{c}3+2i+2j+2l\\
{\Large\Rightarrow}\\{}\end{array}} &|uud,i,j,k+1,l\rangle,\\
|uud,i,j,k,l\rangle &{\footnotesize\begin{array}{c}k\\
{\Large\Rightarrow}\\{}\end{array}} &|uud,i+1,j,k-1,l\rangle,\\
|uud,i,j,k,l\rangle &{\footnotesize\begin{array}{c}i(i+2)\\
{\Large\Rightarrow}\\{}\end{array}} &|uud,i-1,j,k+1,l\rangle,\\
|uud,i,j,k,l\rangle &{\footnotesize\begin{array}{c}k\\
{\Large\Rightarrow}\\{}\end{array}} &|uud,i,j+1,k-1,l\rangle,\\
|uud,i,j,k,l\rangle &{\footnotesize\begin{array}{c}j(j+1)\\
{\Large\Rightarrow}\\{}\end{array}} &|uud,i,j-1,k+1,l\rangle,\\
|uud,i,j,k,l\rangle &{\footnotesize\begin{array}{c}k(1-c_l)^{n-1}\\
{\Large\Rightarrow}\\{}\end{array}} &|uud,i,j,k-1,l+1\rangle,\\
|uud,i,j,k,l\rangle &{\footnotesize\begin{array}{c}l^2\\
{\Large\Rightarrow}\\{}\end{array}} &|uud,i,j,k+1,l-1\rangle,
\end{eqnarray}
and the ``come in" processes as
\begin{eqnarray}
|uud,i,j,k,l\rangle &{\footnotesize\begin{array}{c}3+2i+2j+2l\\
{\Large\Leftarrow}\\{}\end{array}} &|uud,i,j,k-1,l\rangle,\\
|uud,i,j,k,l\rangle &{\footnotesize\begin{array}{c}(3+2i+2j+2l)(k+1)\\
{\Large\Leftarrow}\\{}\end{array}} &|uud,i,j,k+1,l\rangle,\\
|uud,i,j,k,l\rangle &{\footnotesize\begin{array}{c}(i+1)(i+3)\\
{\Large\Leftarrow}\\{}\end{array}} &|uud,i+1,j,k-1,l\rangle,\\
|uud,i,j,k,l\rangle &{\footnotesize\begin{array}{c}k+1\\
{\Large\Leftarrow}\\{}\end{array}} &|uud,i-1,j,k+1,l\rangle,\\
|uud,i,j,k,l\rangle &{\footnotesize\begin{array}{c}(j+1)(j+2)\\
{\Large\Leftarrow}\\{}\end{array}} &|uud,i,j+1,k-1,l\rangle,\\
|uud,i,j,k,l\rangle &{\footnotesize\begin{array}{c}k+1\\
{\Large\Leftarrow}\\{}\end{array}} &|uud,i,j-1,k+1,l\rangle,\\
|uud,i,j,k,l\rangle &{\footnotesize\begin{array}{c}(l+1)(l+1)\\
{\Large\Leftarrow}\\{}\end{array}} &|uud,i,j,k-1,l+1\rangle,\\
|uud,i,j,k,l\rangle &{\footnotesize\begin{array}{c}(k+1)(1-c_{l-1})^{n-2}\\
{\Large\Leftarrow}\\{}\end{array}} &|uud,i,j,k+1,l-1\rangle.
\end{eqnarray}
 From the formula (\ref{balance}), we have the equation
\footnotesize
\begin{eqnarray}
&&\left\{
\left[(3+2i+2j+2l)k\right]+(3+2i+2j+2l)+k+i(i+2)+k+j(j+1)+k(1-c_l)^{n-1}+l^2\right\}
\rho_{i,j,k,l}\nonumber\\ =&&(3+2i+2j +2l
)\rho_{i,j,k-1,l}+\left[(3+2i+2j +2l
)(k+1)\right]\rho_{i,j,k+1,l}+(i+1)(i+3)\rho_{i+1,j,k-1,l}+(k+1)\rho_{i-1,j,k+1,l}\nonumber\\
&&+(j+1)(j+2)\rho_{i,j+1,k-1,l}+(k+1)\rho_{i,j-1,k+1,l}+(l+1)(l+1)\rho_{i,j,k-1,l+1}+(k+1)(1-c_{l-1})^{n-2}\rho_{i,j,k+1,l-1},\nonumber\\
\label{equationgggs}
\end{eqnarray}
\normalsize
where $n=3+2i+2j+2l+k$.

\begin{figure}[htbp]
  \includegraphics{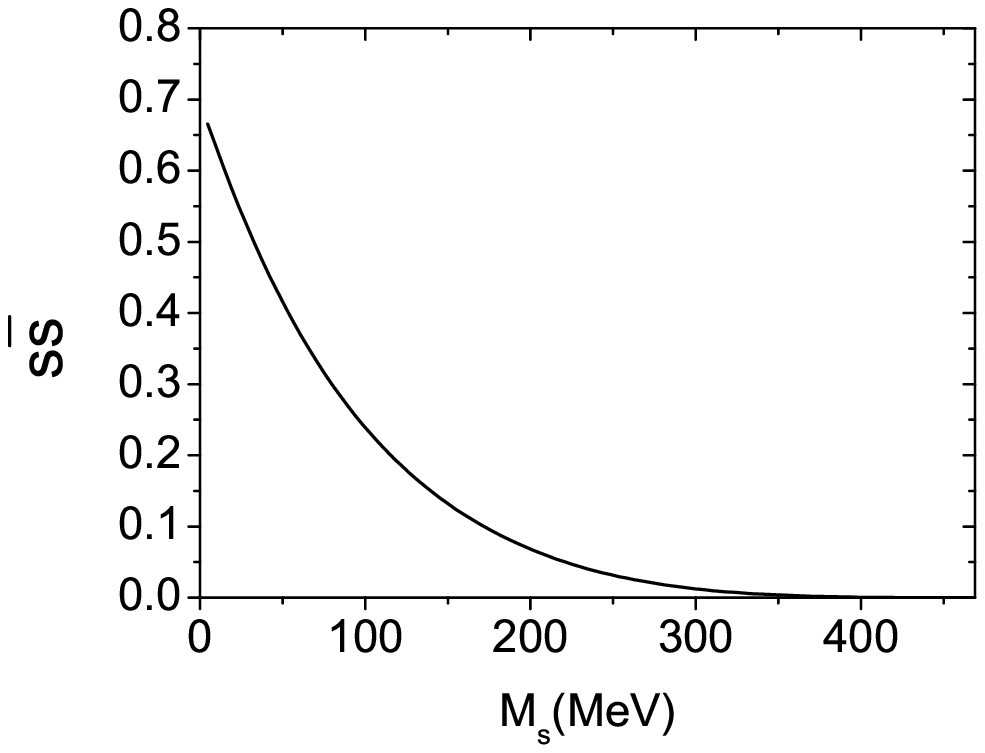}
\caption{The relation between the number of the $s\bar{s}$ pairs
and the strange quark mass $M_s$ in case with the sub-processes
$g\Leftrightarrow gg$ not considered. The number of the $s\bar{s}$
pairs is 0.05 at $M_s\approx 220$MeV. \label{smass-noggg}}
\end{figure}

\begin{figure}[htbp]
  \includegraphics{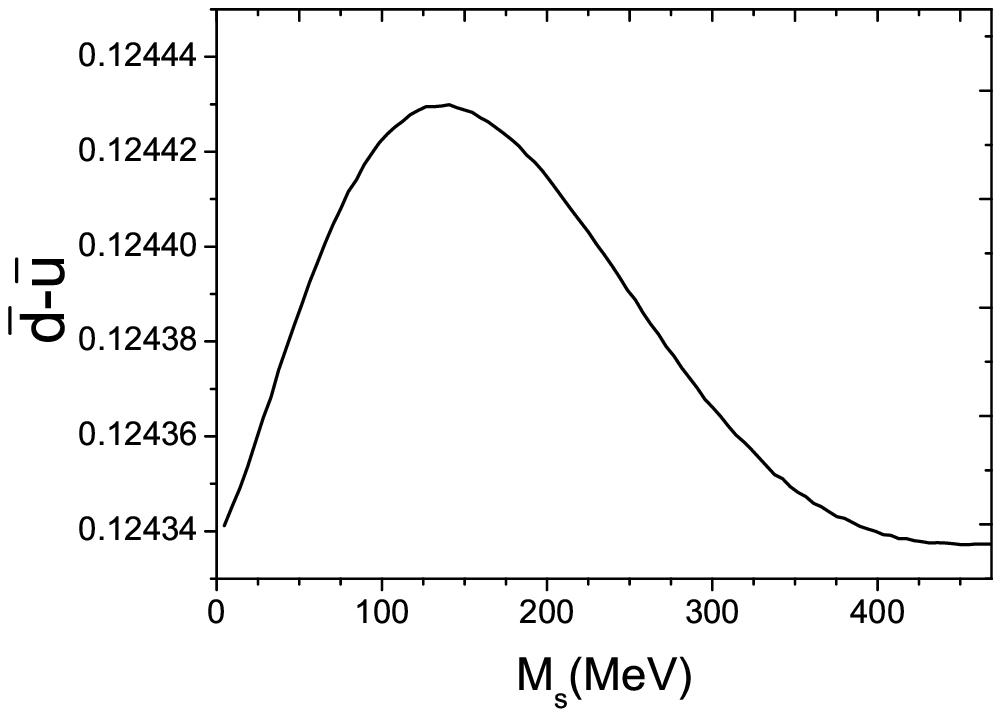}
\caption{The relation between flavor asymmetry $\bar{d}-\bar{u}$
and the strange quark $mass M_s$ in case with the sub-processes
$g\Leftrightarrow gg$ not considered.\label{amass-noggg}}
\end{figure}

Once the strange quark mass $M_s$ is given, we can choose the
upper limits on $i,j,k,l$ to construct an equation set to
determine $\rho_{i,j,k,l}$ and the strange content of proton. We
will take the strange quark mass $M_s$ as a parameter. By solving
the equation set, we give the relation between the number of
$s\bar{s}$ pairs and the strange quark mass $M_s$ in
Fig.~\ref{smass-noggg} and give the relation between
$\bar{d}-\bar{u}$ and $M_s$ in Fig.~\ref{amass-noggg}. When
$0<M_s<\infty$, The parton number's changing scope is
\begin{eqnarray}
&&\bar{u}=\sum_{i,j,k,l}i\rho_{i,j,k,l}\subset[0.309, 0.310],\\
&&\bar{d}=\sum_{i,j,k,l} j\rho_{i,j,k,l}\subset[0.433, 0.434],\\
&&\bar{s}=\sum_{i,j,k,l} l\rho_{i,j,k,l}\subset[0, 0.666],\\ &&
g=\sum_{i,j,k,l} k\rho_{i,j,k,l}\subset[1.000, 1.003],\\ &&
\bar{d}-\bar{u}\subset[0.1243, 0.1245 ].
\end{eqnarray}

\subsection{Strange content with $g\Leftrightarrow gg$
sub-processes considered}

We now consider the strange quark content of the proton with
processes involving the $g\Leftrightarrow gg$ sub-processes also
considered. When using the balance principle to Fock state
$|uud\rangle$, we have the ``go out" process as
\begin{eqnarray}
|uud\rangle{\footnotesize\begin{array}{c}3\\
{\Large\Rightarrow}\\{}\end{array}}|uudg\rangle,
\end{eqnarray}
and the ``come in" process as
\begin{eqnarray}
|uud\rangle{\footnotesize\begin{array}{c}3\\
{\Large\Leftarrow}\\{}\end{array}}|uudg\rangle.
\end{eqnarray}
According to the formula (\ref{balance}), we have the equation
\begin{eqnarray}
3\rho_{uud}=3\rho_{uudg}.
\end{eqnarray}

When using the balance principle to Fock state $|uudg\rangle$, we
have the ``go out" processes as
\begin{eqnarray}
|uudg\rangle &{\footnotesize\begin{array}{c}3\times 1\\
{\Large\Rightarrow}\\{}\end{array}} &|uud\rangle,\\
|uudg\rangle &{\footnotesize\begin{array}{c}3+1\\
{\Large\Rightarrow}\\{}\end{array}} &|uudgg\rangle,\\
|uudg\rangle &{\footnotesize\begin{array}{c}1\\
{\Large\Rightarrow}\\{}\end{array}} &|uud\bar{u}u\rangle,\\
|uudg\rangle &{\footnotesize\begin{array}{c}1\\
{\Large\Rightarrow}\\{}\end{array}} &|uud\bar{d}d\rangle,\\
|uudg\rangle &{\footnotesize\begin{array}{c}1\times(1-c_0)^3\\
{\Large\Rightarrow}\\{}\end{array}} &|uud\bar{s}s\rangle,
\end{eqnarray}
and the ``come in" processes as
\begin{eqnarray}
|uudg\rangle &{\footnotesize\begin{array}{c}3\\
{\Large\Leftarrow}\\{}\end{array}} &|uud\rangle,\\
|uudg\rangle &{\footnotesize\begin{array}{c}3\times 2+1\\
{\Large\Leftarrow}\\{}\end{array}} &|uudgg\rangle,\\
|uudg\rangle &{\footnotesize\begin{array}{c}1\times 3\\
{\Large\Leftarrow}\\{}\end{array}} &|uud\bar{u}u\rangle,\\
|uudg\rangle &{\footnotesize\begin{array}{c}1\times 2\\
{\Large\Leftarrow}\\{}\end{array}} &|uud\bar{d}d\rangle,\\
|uudg\rangle &{\footnotesize\begin{array}{c}1\times 1\\
{\Large\Leftarrow}\\{}\end{array}} &|uud\bar{s}s\rangle.
\end{eqnarray}
Then from the formula (\ref{balance}), we have the equation
\begin{eqnarray}
&&\left[3\times
1+(3+1)+1+1+1\times(1-c_0)^3\right]\rho_{uudg}\nonumber \\
=&&3\rho_{uud}+(3\times 2+1) \rho_{uudgg}+(1\times
3)\rho_{uud\bar{u}u}+(1\times 2)\rho_{uud\bar{d}d}+(1\times
1)\rho_{uud\bar{s}s}.
\end{eqnarray}

Applying the balance principle to a general Fock state
$|uud,i,j,k,l\rangle$, we have the ``go out" processes as
\begin{eqnarray}
|uud,i,j,k,l\rangle &{\footnotesize\begin{array}{c}(3+2i+2j+2l)k+C_k^2\\
{\Large\Rightarrow}\\{}\end{array}} &|uud,i,j,k-1,l\rangle,\\
|uud,i,j,k,l\rangle &{\footnotesize\begin{array}{c}3+2i+2j+2l+k\\
{\Large\Rightarrow}\\{}\end{array}} &|uud,i,j,k+1,l\rangle,\\
|uud,i,j,k,l\rangle &{\footnotesize\begin{array}{c}k\\
{\Large\Rightarrow}\\{}\end{array}} &|uud,i+1,j,k-1,l\rangle,\\
|uud,i,j,k,l\rangle &{\footnotesize\begin{array}{c}i(i+2)\\
{\Large\Rightarrow}\\{}\end{array}} &|uud,i-1,j,k+1,l\rangle,\\
|uud,i,j,k,l\rangle &{\footnotesize\begin{array}{c}k\\
{\Large\Rightarrow}\\{}\end{array}} &|uud,i,j+1,k-1,l\rangle,\\
|uud,i,j,k,l\rangle &{\footnotesize\begin{array}{c}j(j+1)\\
{\Large\Rightarrow}\\{}\end{array}} &|uud,i,j-1,k+1,l\rangle,\\
|uud,i,j,k,l\rangle &{\footnotesize\begin{array}{c}k(1-c_l)^{n-1}\\
{\Large\Rightarrow}\\{}\end{array}} &|uud,i,j,k-1,l+1\rangle,\\
|uud,i,j,k,l\rangle &{\footnotesize\begin{array}{c}l^2\\
{\Large\Rightarrow}\\{}\end{array}} &|uud,i,j,k+1,l-1\rangle,
\end{eqnarray}
and the ``come in" processes as
\begin{eqnarray}
|uud,i,j,k,l\rangle &{\footnotesize\begin{array}{c}3+2i+2j+2l+k-1\\
{\Large\Leftarrow}\\{}\end{array}} &|uud,i,j,k-1,l\rangle,\\
|uud,i,j,k,l\rangle &{\footnotesize\begin{array}{c}(3+2i+2j+2l)(k+1)+C_{k+1}^2\\
{\Large\Leftarrow}\\{}\end{array}} &|uud,i,j,k+1,l\rangle,\\
|uud,i,j,k,l\rangle &{\footnotesize\begin{array}{c}(i+1)(i+3)\\
{\Large\Leftarrow}\\{}\end{array}} &|uud,i+1,j,k-1,l\rangle,\\
|uud,i,j,k,l\rangle &{\footnotesize\begin{array}{c}k+1\\
{\Large\Leftarrow}\\{}\end{array}} &|uud,i-1,j,k+1,l\rangle,\\
|uud,i,j,k,l\rangle &{\footnotesize\begin{array}{c}(j+1)(j+2)\\
{\Large\Leftarrow}\\{}\end{array}} &|uud,i,j+1,k-1,l\rangle,\\
|uud,i,j,k,l\rangle &{\footnotesize\begin{array}{c}k+1\\
{\Large\Leftarrow}\\{}\end{array}} &|uud,i,j-1,k+1,l\rangle,\\
|uud,i,j,k,l\rangle &{\footnotesize\begin{array}{c}(l+1)(l+1)\\
{\Large\Leftarrow}\\{}\end{array}} &|uud,i,j,k-1,l+1\rangle,\\
|uud,i,j,k,l\rangle &{\footnotesize\begin{array}{c}(k+1)(1-c_{l-1})^{n-2}\\
{\Large\Leftarrow}\\{}\end{array}} &|uud,i,j,k+1,l-1\rangle.
\end{eqnarray}
Thus we have the equation \footnotesize
\begin{eqnarray}
&&\left\{
\left[(3+2i+2j+2l)k+C_k^2\right]+(3+2i+2j+2l+k)+k+i(i+2)+k+j(j+1)+k(1-c_l)^{n-1}+l^2\right\}
\rho_{i,j,k,l}\nonumber\\
=&&(3+2i+2j+2l+k-1)\rho_{i,j,k-1,l}+\left[(3+2i+2j+2l)(k+1)+C_{k+1}^2\right]
\rho_{i,j,k+1,l}+(i+1)(i+3)\rho_{i+1,j,k-1,l}\nonumber\\
&&+(k+1)\rho_{i-1,j,k+1,l}+(j+1)(j+2)\rho_{i,j+1,k-1,l}+(k+1)\rho_{i,j-1,k+1
,l } +(l+1)(l+1)\rho_{i,j,k-1,l+1}\nonumber \\
&&+(k+1)(1-c_{l-1})^{n-2}\rho_{i,j,k+1,l-1} \label{equationgggs2}
\end{eqnarray}
\normalsize
where $n=3+2i+2j+2l+k$.

\begin{figure}[htbp]
  \includegraphics{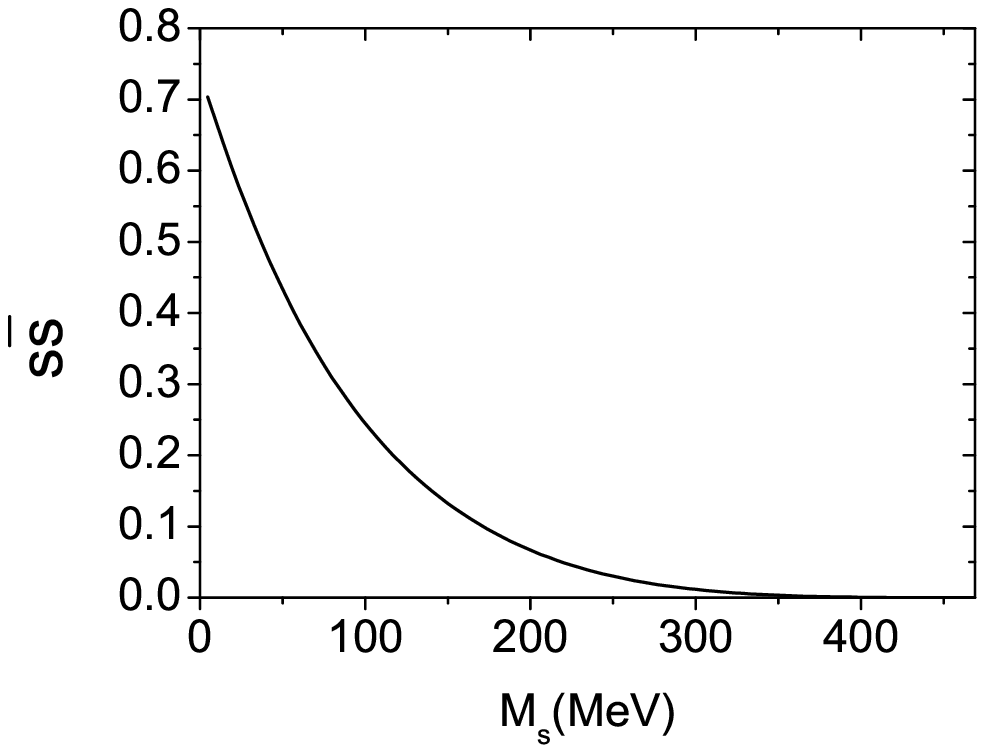}
\caption{The relation between the number of the $s\bar{s}$ pairs
and the strange quark mass $M_s$ in case with the sub-processes
$g\Leftrightarrow gg$ considered. The number of the $s\bar{s}$
pairs is 0.05 at $M_s\approx 220$MeV.\label{smass}}
\end{figure}

\begin{figure}[htbp]
 \includegraphics{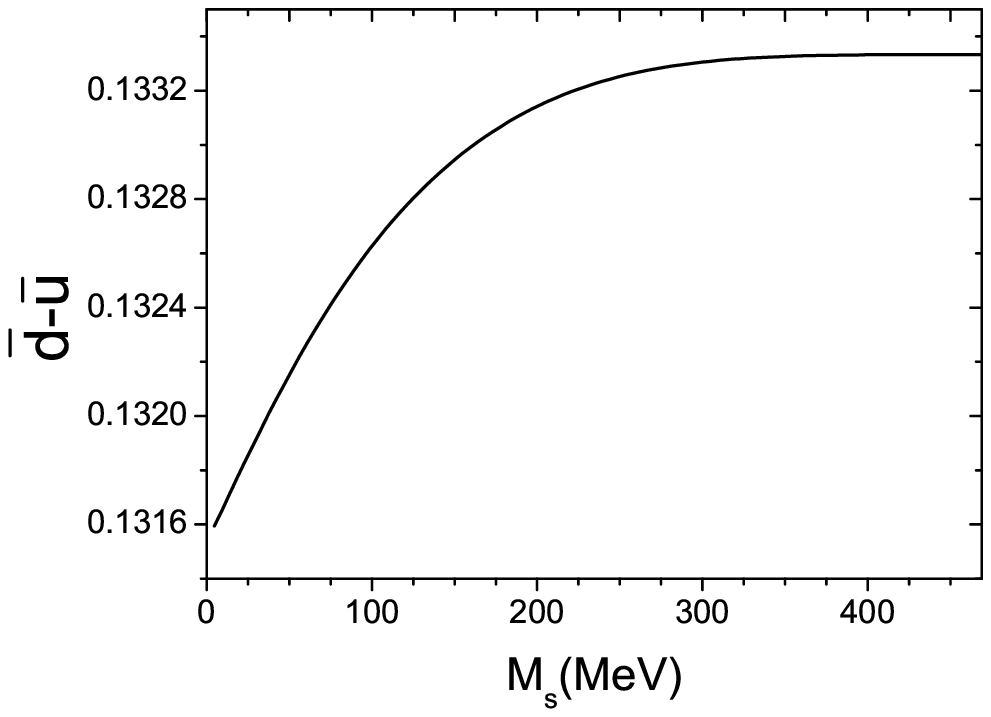}
\caption{The relation between flavor asymmetry $\bar{d}-\bar{u}$
and the strange quark mass $M_s$ in case with the sub-processes
$g\Leftrightarrow gg$ considered.\label{amass}}
\end{figure}

By solving the equation set constructed from equation
(\ref{equationgggs2}) after choosing the upper limits on
$i,j,k,l$, we give the relation between the number of $s\bar{s}$
pairs and the strange quark mass $M_s$ in Fig.~\ref{smass} and
give the relation between $\bar{d}-\bar{u}$ and $M_s$ in
Fig.~\ref{amass}. When $0<M_s<\infty$, the parton number's
changing scope is
\begin{eqnarray}
&&\bar{u}=\sum_{i,j,k,l}i\rho_{i,j,k,l}\subset[0.331, 0.337],\\
&&\bar{d}=\sum_{i,j,k,l} j\rho_{i,j,k,l}\subset[0.463, 0.470],\\
&&\bar{s}=\sum_{i,j,k,l} l\rho_{i,j,k,l}\subset[0, 0.703],\\ &&
g=\sum_{i,j,k,l} k\rho_{i,j,k,l}\subset[1.079, 1.099],\\ &&
\bar{d}-\bar{u}\subset[0.131, 0.134 ].
\end{eqnarray}
We find that the number of $s\bar{s}$ pairs equals to 0.05 at $M_s
\approx 220$~MeV, which is in the reasonable range of the strange
mass. We also notice that the introduction of the strange sea
content brings very small influence to the results for the
light-flavor sea content.

\section{conclusion}

We improved the detailed balance model to a balance model because
the detailed balance model leads to inconsistency some times. We
find that there is about 10 percent difference between the results
of these two models in case with sub-processes $g\Leftrightarrow
gg$ also considered. The detailed balance model is simple and
perspicuous, and it is a good approximation of the balance model,
while the balance model is from a more general consideration, but
is complicated, and needs to introduce the upper limits on
$i,j,k,l$ and to solve an equation set with a big number of
equations. To consider the strange content inside the proton, we
use the equal probability assumption to get the gluon energy
distribution and from which, we get the relation between the
suppressing factor of $g\rightarrow s\bar{s}$ and the strange
quark mass. Then in the balance model, the number of $s\bar{s}$
pairs in the proton can be determined once the strange quark mass
is given.

We make some comments about the limitations of our results such as
those in Table I. The quarks and gluons in the Fock states are the
``intrinsic" partons of the proton, since they are multi-connected
non-perturbatively to the valence quarks \cite{Bro96}. Such
partons are different from the ``extrinsic" partons generated from
the QCD hard bremsstrahlung and gluon-splitting as part of the
lepton scattering interaction. Thus the results in our paper are
expected to work at a scale for the ``intrinsic" partons, which is
about $Q^2 \sim 1$ GeV$^2$. The total number of intrinsic partons
inside the proton is around 5.5, which is a small number of
particles for the feasibility of the statistical method.

The detailed balance model and balance model can be taken as a new
statistical model besides other successful statistical models such
as \cite{Bhalerao} and \cite{Soffer}. We also noticed that the
effect from the interaction of the quark-antiquark pairs with the
quarks inside the nucleon has been also considered in a simple
model that can account for the light-flavor sea asymmetry
\cite{DiSalvo}.  The advantage of the balance model is that a
complete set of all quark-gluon Fock states can be obtained
without any parameter in the case with only light-flavor quarks,
but we still need to deal with the more complicated situation with
parton polarizations taken into account for more general
applications.

{\bf Acknowledgments:} We are very grateful to Prof.~Li-Ming Yang
for his supervision and encouragement. This work is partially
supported by National Natural Science Foundation of China.

\vspace{5cm}

\mediumtext
\begin{table}
  \caption{The probabilities, $\rho_{i,j,k}$,
  of finding the quark-gluon Fock states of the
  proton, calculated using the principle of balance without any parameter.
   $|uud,i,j,k\rangle$
   is the Fock state of proton, $i$ is the number of $u\bar{u}$ quark pairs,
   $j$ is the number
   of $d\bar{d}$ pairs, and $k$ is the number of gluons. }
        \label{table}
        \begin{tabular}{cclcccccc}
        {i}&{j}&{$|uud,i,j,0\rangle$}&{$\rho_{i,j,0}$}&{$\rho_{i,j,1}$}&{$\rho_{i,j,2}$}&{$\rho_{i,j,3}$}
        &{$\rho_{i,j,4}$}&{$\cdots$}\\
    \tableline
        0&0&$|uud\rangle$         &0.148793&0.148793&0.081887&0.032056&0.009876&$\cdots$\\
        1&0&$|uud\bar{u}u\rangle$ &0.052960&0.054978&0.030419&0.011787&0.003564&$\cdots$\\
        0&1&$|uud\bar{d}d\rangle$ &0.080334&0.082708&0.045441&0.017500&0.005263&$\cdots$\\
        1&1&$|uud\bar{u}u\bar{d}d\rangle$ &0.029306&0.030569&0.016763&0.006386&0.001889&$\cdots$\\
        0&2&$|uud\bar{d}d\bar{d}d\rangle$ &0.014570&0.015242&0.008381&0.003201&0.000949&$\cdots$\\
        2&0&$|uud\bar{u}u\bar{u}u\rangle$ &0.007231&0.007642&0.004240&0.001632&0.000488&$\cdots$\\
        1&2&$|uud\bar{u}u\bar{d}d\bar{d}d\rangle$ &0.005355&0.005620&0.003073&0.001160&0.000339&$\cdots$\\
        2&1&$|uud\bar{u}u\bar{u}u\bar{d}d\rangle$ &0.004011&0.004223&0.002315&0.000877&0.000257&$\cdots$\\
        0&3&$|uud\bar{d}d\bar{d}d\bar{d}d\rangle$ &0.001327&0.001403&0.000773&0.000294&0.000086&$\cdots$\\
        3&0&$|uud\bar{u}u\bar{u}u\bar{u}u\rangle$ &0.000531&0.000567&0.000315&0.000121&0.000036&$\cdots$\\
         $\cdots$&$\cdots$&$\cdots$&$\cdots$&$\cdots$&$\cdots$&$\cdots$&$\cdots$&$\ddots$\\
         \end{tabular}
   \end{table}

\end{document}